\documentclass[aps,nofootinbib, showpacs,
]{revtex4}
\usepackage{graphicx}
\usepackage{amsmath}
\usepackage{amsfonts}
\usepackage{amssymb}
\usepackage{color}

\newcommand{\omits}[1]{}

\begin{document}


\title{Spontaneous Pair Production in Reissner-Nordstr{\"o}m Black Holes}

\author{Chiang-Mei Chen$^1$} \email{cmchen@phy.ncu.edu.tw}
\author{Sang Pyo Kim$^{2,4,5}$} \email{sangkim@kunsan.ac.kr}
\author{I-Chieh Lin$^1$} \email{wcid0o@yahoo.com.tw}
\author{Jia-Rui Sun$^{3,1}$} \email{jrsun@ecust.edu.cn}
\author{Ming-Fan Wu$^1$} \email{adiemuswu@gmail.com}

\affiliation{${}^1$Department of Physics and Center for Mathematics and Theoretical Physics, National Central University, Chungli 320, Taiwan}
\affiliation{${}^2$Department of Physics, Kunsan National University, Kunsan 573-701, Korea}
\affiliation{${}^3$Department of Physics and Institute of Modern Physics, East China University of Science and Technology, Shanghai 200237, China}
\affiliation{${}^4$Institute of Astrophysics, Center for Theoretical Physics, Department of Physics, National Taiwan University, Taipei 106, Taiwan}
\affiliation{${}^5$Yukawa Institute for Theoretical Physics, Kyoto University, Kyoto 606-8502, Japan}


\begin{abstract}
We investigate the spontaneous pair production, including the Schwinger mechanism and the Hawking thermal radiation, of charged scalar particles from the near horizon region of a (near) extremal Reissner-Nordstr\"om black hole. The paradigm is equivalent to the dynamics of the charged scalar field in a specific AdS$_2 \times S^2$ spacetime with a constant electric field. Two possible boundary conditions are adopted to explicitly compute the corresponding production rate and absorption cross section. It is shown that the Schwinger production rate can be eventually suppressed by the increasing attractive gravitational force as the geometry changes from the extremal to the near extremal black hole. Consequently, the Schwinger mechanism and the Hawking radiation are generically indistinguishable for the near extremal black holes. The holographic description dual to the pair production is also briefly discussed.
\end{abstract}

\pacs{04.62.+v, 04.70.Dy, 12.20.-m}

\maketitle

\section{Introduction}
The spontaneous pair production in a strong background field or in a causally disconnected spacetime is a significant and profound quantum effect. The Schwinger mechanism and the Hawking radiation are such a phenomenon in which virtual pairs of particles and antiparticles from vacuum fluctuations are separated into real pairs either by an electric field~\cite{Schwinger:1951nm} or by a black hole horizon via a tunneling~\cite{Parikh:1999mf}. These two processes, though independent of each other, could be generically intertwined and indistinguishable in the background of a charged black hole.
Besides, the emission of charged pairs affects the charge to mass ratio of the Reissner-Nordstr\"om (RN) black hole~\cite{Zaumen:1974} and this effect is more rapid than the loss of mass and angular momentum~\cite{Carter:1974yx, Damour:1974qv}. It has been shown in Ref.~\cite{Gibbons:1975kk} that for a large RN black hole with a small Hawking temperature, the vacuum polarization by a charged scalar field is dominated by the Schwinger mechanism. The Hawking emission of spin-1/2 fermions~\cite{Page:1977um} from and the evolution~\cite{Hiscock:1990ex} of an RN black hole have been numerically studied. Also it has been shown that the semiclassical tunneling probability of charged particles in the RN black hole leads to the Schwinger formula on the black hole horizon~\cite{Khriplovich:1999gm}. For more recent related discussions about the quantum effect of fields in the charged black hole backgrounds, see Refs.~\cite{Gabriel:2000mg, Belgiorno:2007va, Belgiorno:2008mx, Belgiorno:2009pq}.

The spontaneous pair production from a black hole is expected to occur mainly at the near horizon region: the causal boundary for the Hawking radiation and electric field domination for the Schwinger mechanism. This expectation will be confirmed by the fact that in the leading order the pair production rate agrees with the Schwinger formula derived in the whole RN spacetime. In this paper we analytically study the emission from an RN black hole by observing that the spacetime of the near horizon region of the (near) extremal RN black hole has a particular product structure AdS$_2 \times S^2$ with the same radius. Thus, the positive curvature of the $S^2$ exactly compensates the negative part of the AdS$_2$. Moreover, the electric field is constant and determined by the geometrical radius. This background provides a simple framework in which the analysis of spontaneous pair production can be exactly performed. We consider the dynamics of a charged scalar field in a (near) extremal RN black hole. The key equation of motion is very analogous to the one studied in Ref.~\cite{Kim:2008xv} for an AdS$_2$ spacetime. The additional integrants include a standard separation constant for spherical harmonics on the $S^2$-section and a parameter associated with the black hole temperature. It turns out that the process of spontaneous production corresponds to the instability of the charged scalar in AdS$_2$, namely, the violation of the Breitenlohner-Freedman (BF) bound~\cite{Breitenlohner:1982jf, Breitenlohner:1982bm} which results the complex conformal weight of the operators dual to the charged scalar field in the AdS/CFT correspondence~\cite{Maldacena:1997re, Gubser:1998bc, Witten:1998qj}.

A direct approach to obtain the physical information, such as the mean number of pairs and the absorption cross section, about the spontaneous production is to compute the ratios of fluxes passing through each boundary under a suitable boundary condition. The Bogoliubov coefficients can be found by applying the same boundary condition that resolves the Klein paradox for the tunneling barrier in static configurations of electric fields in quantum electrodynamics~\cite{Nikishov:1970br, Damour:1977, Hansen:1980nc} (see also Ref.~\cite{Kim:2011dn}). In fact, there are two boundaries in the near horizon geometry of the (near) extremal RN black holes: an outer boundary at ``spatial infinity'' of the near horizon region and an inner boundary associated to the black hole causal horizon. Each of following boundary conditions admits to compute the desired physical quantities: no flux flowing into the considered spacetime via either the outer boundary or the inner boundary. The physical interpretations of these two boundary conditions will be discussed in Section~\ref{SBC}.

Under these two boundary conditions, the flux either incoming from the asymptotic boundary or outgoing from the horizon\footnote{Our classification of in and out states is with respect to the view point of black hole. Thus, the outgoing flux is positive and incoming flux is negative.} has been turned off. Therefore, there are only three non-vanishing fluxes, naturally named as incident, reflected and transmitted. Moreover, the flux conservation reduces the number of independent fluxes down to two. The different ratios of fluxes give various physically interesting quantities: in particular, the magnitude of the vacuum persistence amplitude $|\alpha|^2$, the mean number of pairs $|\beta|^2$, and the absorption cross section (grey body factor) $\sigma_\mathrm{abs}$. Remarkably, the mean number of pairs and the absorption cross section are the same for the both boundary conditions. This equivalence is a consequence of the unitarity of the scattering matrix for a given quantum number of the charged scalar field, which reflects the flux conservations~\cite{Merzbacher:1970}.

We explicitly compute these three quantities for the extremal and near extremal RN black holes. We perform the calculations in the both outer and inner boundary conditions to confirm the equivalence.  Obviously but interestingly enough, the extremal RN black hole has the zero Hawking temperature, and thus the pair production is completely generated by the Schwinger mechanism. However, for the near extremal RN black hole, though the Hawking thermal radiation is included, the production rate is indeed reduced. The decreasing production rate is due to the increasing ``attractive'' gravitational force as the geometry changes from the extremal to the near extremal RN black hole, which suppresses the Schwinger mechanism. Thus this result implies that the Schwinger mechanism and the Hawking radiation are generically mixed and cannot be distinguished simply by imposing different boundary conditions. In addition, we also compute the corresponding quasi-normal modes for the charged scalar field by imposing both the inner and outer boundary conditions at the same time and and show that there is no quasi-normal modes in the extremal limit. We further discuss the dual CFT description of the pair production process and find that the absorption cross section of the charged scalar field calculated from the gravity side matches with that of its dual operator in the 2D CFT based on the RN/CFT correspondence studied in~\cite{Chen:2010bsa, Chen:2010as, Chen:2011gz}.

The outline of this paper is as follows. Firstly, we analyze the dynamics of a charged scalar field in the near horizon region of (near) extremal RN black holes. In Section III we discuss the outer and inner boundary conditions and show the equivalence of the mean number of pairs and the absorption cross section from the both boundary conditions. The results of spontaneous production are presented in Section IV for the extremal case and in Section V for the near extremal case. Further, a holographic interpretation is given in Section VI. Finally, we conclude our results in Section VII. In Appendix A we summarize the properties of the special functions that are used in this paper. A comparison, using the Hamilton-Jacobi equation and the phase-integral method, is made with previously known results in Appendix B.

\section{Scalar Field in the RN black hole}
The Reissner-Nordstr{\"o}m (RN) solution of a charged black hole with the following metric and a $U(1)$ gauge field, $F = dA$, [in units of $c = \hbar = G = 1$]
\begin{eqnarray} \label{RN}
ds^2 &=& - \left( 1 - \frac{2 M}{r} + \frac{Q^2}{r^2} \right) dt^2 + \frac{dr^2}{1 - \frac{2 M}{r} + \frac{Q^2}{r^2}} + r^2 d\Omega_2^2,
\nonumber\\
A &=& \frac{Q}{r} dt; \qquad F = \frac{Q}{r^2} dt \wedge dr,
\end{eqnarray}
is characterized by the mass $M$ and the charge $Q$. Here $d\Omega_2^2 = d\theta^2 + \sin^2\theta d\phi^2$ denotes the metric on a unit two-sphere. By taking the near horizon and the near-extremal limits, together with a suitable rescaling of the time
\begin{equation} \label{rescal}
r \to Q + \epsilon \rho, \qquad M \to Q + \frac{\epsilon^2 B^2}{2 Q}, \qquad t \to \frac{\tau}{\epsilon},
\end{equation}
one can show that the near horizon region of a near extremal RN black hole corresponds a black hole like solution as
\begin{eqnarray} \label{NHRN}
ds^2 &=& - \frac{\rho^2 - B^2}{Q^2} d\tau^2 + \frac{Q^2}{\rho^2 - B^2} d\rho^2 + Q^2 d\Omega_2^2,
\nonumber\\
A &=& -\frac{\rho}{Q} d\tau; \qquad F = \frac{1}{Q} d\tau \wedge d\rho.
\end{eqnarray}
The parameter $B$ labels the ``rescaled'' deviation~(\ref{rescal}) from the extremal limit and acts as the horizon radius of the new black hole solution. Noting that the metric~(\ref{NHRN}) has an AdS$_2 \times S^2$ geometry with the same radius of $Q$ and the strength of the associated gauge field is constant with the magnitude $Q$, we will study the spontaneous pair production of a charged scalar field in this background.

The action for a probe charged scalar field $\Phi$ with the mass $m$ and the charge $q$ is
\begin{equation} \label{action}
S = \int d^4x \sqrt{-g} \left( - \frac12 D_\alpha \Phi^* D^\alpha \Phi - \frac12 m^2 \Phi^*\Phi \right),
\end{equation}
where $D_{\alpha} \equiv \nabla_{\alpha} - i q A_{\alpha}$ with $\nabla_\alpha$ being the covariant derivative in curved spacetime. The corresponding Klein-Gordon (KG) equation
\begin{equation} \label{eom}
(\nabla_\alpha - i q A_\alpha) (\nabla^\alpha - i q A^\alpha) \Phi - m^2 \Phi = 0
\end{equation}
has the flux of a probe charged scalar field
\begin{equation} \label{flux}
D = i \sqrt{-g} g^{\rho\rho} (\Phi D_\rho \Phi^* - \Phi^* D_\rho \Phi),
\end{equation}
which is positive for an outgoing mode and is negative for an ingoing mode.

We look for the solution of the scalar field
\begin{equation} \label{ansatz}
\Phi(\tau, \rho, \theta, \phi) = \mathrm{e}^{-i \omega \tau + i n \phi} R(\rho) S(\theta),
\end{equation}
which separates the KG equation as
\begin{eqnarray}
\partial_\rho \left[ (\rho^2 - B^2) \partial_\rho R \right] + \left[ \frac{(q \rho - \omega Q)^2 Q^2}{\rho^2 - B^2} - m^2 Q^2 - \lambda_l \right] R &=& 0, \label{EqR}
\\
\frac1{\sin\theta} \partial_\theta (\sin\theta \partial_\theta S) - \left( \frac{n^2}{\sin^2\theta} - \lambda_l \right) S &=& 0,
\end{eqnarray}
where $\lambda_l$ is a separation constant. The solution for $S(\theta)$ is the standard spherical harmonics with the eigenvalue $\lambda_l = l (l + 1)$.

The radial equation~(\ref{EqR}) can be understood as the equation of motion of a probe scalar field $R(\rho)$ with an effective mass $m_\mathrm{eff}^2 = m^2 - q^2 + \lambda_l/Q^2$ propagating in an AdS$_2$ geometry\footnote{Notice that Eq.~(\ref{EqR}) can also be interpreted as the equation of motion of a massive scalar field with effective mass $\tilde{m}_\mathrm{eff}^2 = 4 m_\mathrm{eff}^2$ propagating in an ``effective'' AdS$_3 = $ AdS$_2 \times S^1$ background, by treating the U(1) gauge field as the additional $S^1$ fiber over the geometrical AdS$_2$, from the viewpoint of the hidden conformal symmetry~\cite{Chen:2010ywa, Chen:2012tj}. The effective AdS$_3$ structure has the same radius $L_\mathrm{AdS} = Q$ as the AdS$_2$ case, and then the condition~(\ref{BFv}) that is required for the existence of propagating modes is actually
\begin{equation} \label{vBFads3}
\tilde{m}_\mathrm{eff}^2 = 4 m_\mathrm{eff}^2 = 4 \left( m^2 - q^2 + \frac{\lambda_l}{Q^2} \right) < - \frac1{Q^2},
\end{equation}
which violates the BF bound in the AdS$_3$ spacetime.} with the radius $L_\mathrm{AdS} = Q$. It is well known that
an instability will occur when the effective mass square is less than the Breitenlohner-Freedman (BF) bound~\cite{Breitenlohner:1982jf, Breitenlohner:1982bm}, more precisely in a general AdS$_{d+1}$ space of radius $L_\mathrm{AdS}$
\begin{equation} \label{BFb}
m_\mathrm{eff}^2 < - \frac{d^2}{4 L_\mathrm{AdS}^2}.
\end{equation}
Therefore the violation of the BF bound in the AdS$_2$ spacetime to ensure the presence of the Schwinger pair production and/or the Hawking radiation~\cite{Pioline:2005pf} requires
\begin{equation}\label{BFv}
(m^2 - q^2) Q^2 + \left( l + \frac12 \right)^2 < 0.
\end{equation}
This condition implies that the mass of a created particle (which is an unstable tachyon mode) should be smaller than its charge, and the background electric field should be larger than a threshold value. Thus, a neutral scalar particle, i.e. $q = 0$, cannot be produced by the Hawking radiation in the near extremal case. Also the energetic condition for pair production requires $q Q/r_H > m $. In the near extremal black hole $r_H \approx Q$, pairs are produced when $q > m$, which is satisfied by electrons and positrons.

\section{Boundary Conditions and Their Equivalence} \label{SBC}

\subsection{Outer Boundary Condition}
\begin{figure}[ht]
\includegraphics[width=10cm]{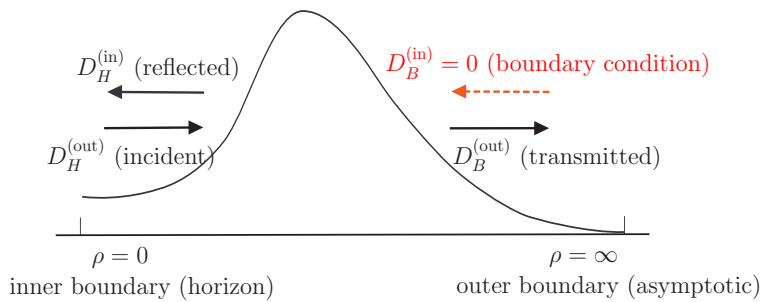}
\caption{The outer boundary condition: no incoming flux at the asymptotic outer boundary.} \label{F1}
\end{figure}

One possible boundary condition to reveal the behavior of the spontaneous pair production is to require no incoming flux at the asymptotic outer boundary, see Fig.~\ref{F1}. In the St\"{u}ckelberg-Feynman picture, the outgoing (transmitted) flux at the asymptotic represents the spontaneously produced ``particle'' while the outgoing (incident) flux at the inner boundary can be interpreted as the total particles created by vacuum fluctuations, and the incoming (reflected) flux represents the portion that was re-annihilated.

In such an intuitive picture, the flux conservation
\begin{equation}
| D_\mathrm{incident} | = | D_\mathrm{reflected} | + | D_\mathrm{transmitted} |,
\end{equation}
leads to the Bogoliubov relation
\begin{equation}
|\alpha|^2 - |\beta|^2 = 1,
\end{equation}
where the vacuum persistence amplitude $|\alpha|^2$ and the mean number of produced pairs $|\beta|^2$ from the Bogoliubov coefficients $\alpha$ and $\beta$ are given by the ratios of the incident and the transmitted coefficients to the reflected one in the Coulomb gauge (time-independent gauge potential)~\cite{Kim:2000un, Kim:2009pg}
\begin{equation}
|\alpha|^2 \equiv \frac{D_\mathrm{incident}}{D_\mathrm{reflected}}, \qquad |\beta|^2 \equiv \frac{D_\mathrm{transmitted}}{D_\mathrm{reflected}}.
\end{equation}

\subsection{Inner Boundary Condition}
\begin{figure}[ht]
\includegraphics[width=10cm]{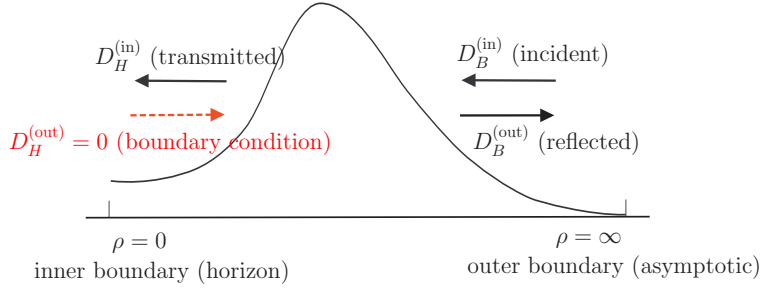}
\caption{The inner boundary condition: no outgoing flux at the inner boundary.} \label{F2}
\end{figure}

The alternative boundary condition is to require no outgoing flux at the inner boundary (horizon), see Fig.~\ref{F2}. In this case, the incoming (transmitted) flux at the horizon can be understood as the spontaneous produced ``antiparticles''. Similarly, the incoming (incident) and outgoing (reflected) fluxes at the asymptotic boundary can be interpreted again as the total created antiparticles and the re-annihilated part.

Moreover, this boundary condition can have another intuitive viewpoint as a scattering process of an incident flux coming from the asymptotic boundary. In addition to $|\alpha|^2$ and $|\beta|^2$, it is also useful to define the absorption cross section as
\begin{equation}
\sigma_\mathrm{abs} \equiv \frac{D_\mathrm{transmitted}}{D_\mathrm{incident}} = \frac{|\beta|^2}{|\alpha|^2}.
\end{equation}

\subsection{Equivalence} \label{equivalence}
According to a naive physical picture of the outer and inner boundary conditions, it is natural to expect that these two boundary conditions indeed are equivalent since the particles and antiparticles should always appear in pairs due to the charge conservation and/or the energy-momentum conservation, no matter whether they are created or separated. Note that the flux, $D$, at each boundary is the magnitude square of the coefficient, ${\cal D}$, for the incoming wave or the outgoing wave times the corresponding momentum, i.e. $D = |{\cal D}|^2$. Actually, these coefficients are related by the scattering matrix as~\cite{Merzbacher:1970}
\begin{equation} \label{s-matrix}
\begin{pmatrix} {\cal D}_H^\mathrm{(in)} \\ {\cal D}_B^\mathrm{(out)} \end{pmatrix} =
\begin{pmatrix} {\cal S}_{11} & {\cal S}_{12} \\ {\cal S}_{21} & {\cal S}_{22} \end{pmatrix}
\begin{pmatrix} {\cal D}_H^\mathrm{(out)} \\ {\cal D}_B^\mathrm{(in)} \end{pmatrix}.
\end{equation}
The unitarity of scattering matrix, ${\cal S}^{\dagger} {\cal S} = {\cal S} {\cal S}^{\dagger} =1$, leads to
\begin{eqnarray} \label{unit-con}
|{\cal S}_{11}| = |{\cal S}_{22}|, &\quad& |{\cal S}_{12}| = |{\cal S}_{21}|,
\nonumber\\
|{\cal S}_{11}|^2 + |{\cal S}_{12}|^2 = 1, &\quad& {\cal S}_{11} {\cal S}_{12}^* + {\cal S}_{21} {\cal S}_{22}^* = 0.
\end{eqnarray}
The mean numbers in these two different boundary conditions
\begin{eqnarray}
\left| \beta_\mathrm{outer} \right|^2 &=& \frac{D_B^\mathrm{(out)}}{D_H^\mathrm{(in)}} \Biggr|_{D_B^\mathrm{(in)} = 0} = \left| \frac{{\cal S}_{21}}{{\cal S}_{11}} \right|^2,
\nonumber\\
\left| \beta_\mathrm{inner} \right|^2 &=& \frac{D_H^\mathrm{(in)}}{D_B^\mathrm{(out)}} \Biggr|_{D_H^\mathrm{(out)} = 0} = \left| \frac{{\cal S}_{12}}{{\cal S}_{22}} \right|^2,
\end{eqnarray}
are thus the same due to Eq.~(\ref{unit-con}). Then the equivalence between the absorption cross sections follows from the Bogoliubov relation $|\alpha|^2 - |\beta|^2 = 1$:
\begin{equation} \label{betasigma}
\sigma_\mathrm{abs} = \frac{|\beta|^2}{|\alpha|^2} = \frac{|\beta|^2}{1 + |\beta|^2}.
\end{equation}

\section{Extremal Black Holes}

\subsection{Solution of Scalar Field}
The simplest background to study the Schwinger pair production is the case $B = 0$ in which the Hawking temperature vanishes and  therefore no thermal radiations exist. By defining a new coordinate
\begin{equation}
z \equiv i \frac{2 \omega Q^2}{\rho}, \qquad \arg z = \frac{\pi}2,
\end{equation}
the radial equation~(\ref{EqR}) can be expressed as the Whittaker equation
\begin{equation}
\left( \partial^2_z - \frac14 + \frac{i a}{z} + \frac{\frac14 - (i b)^2}{z^2} \right) R(z) = 0,
\end{equation}
where the parameters $a$ and $b$ are defined as
\begin{equation} \label{ab}
a \equiv qQ, \qquad b \equiv \sqrt{(q^2 - m^2) Q^2 - \left( l + \frac12 \right)^2}.
\end{equation}
The BF bound violation~(\ref{BFv}) requires the parameter $b$ to be real. The general solution for $R(\rho)$ can be expressed in terms of the Whittaker function $M_{\kappa, \mu}(z)$ as
\begin{equation} \label{SolR}
R(z) = c_1 M_{ia, -ib}(z) + c_2 M_{ia, ib}(z).
\end{equation}

In order to study the pair production rate, one needs to distinguish the in- and out-states at the both inner and outer boundaries. For this purpose we first analyze the behavior of the solution~(\ref{SolR}) at the asymptotic boundary and on the horizon, respectively. In the asymptotic limit, $\rho \to \infty$ or $z \to 0$, by using the property~(\ref{limMW}), the solution~(\ref{SolR}) at the outer boundary approaches to
\begin{equation} \label{Rasy}
R_B(z) = c_B^\mathrm{(in)} \, z^{\frac12 + ib} + c_B^\mathrm{(out)} \, z^{\frac12 - ib},
\end{equation}
where
\begin{equation} \label{cB}
c_B^\mathrm{(in)} = c_1, \qquad c_B^\mathrm{(out)} = c_2.
\end{equation}
In this case, the factor $z^{ib} \sim \rho^{-ib}$ represents the ingoing mode while $z^{-ib} \sim \rho^{ib}$ represents the outgoing mode. In the near horizon limit, $\rho \to 0$ or $z \to \infty$, by applying the properties~(\ref{relM}, \ref{limMW}) the solution reduces to
\begin{equation}
R_H(z) =  c_H^\mathrm{(in)} e^{\frac{z}2} z^{-ia} + c_H^\mathrm{(out)} e^{-\frac{z}2} z^{ia},
\end{equation}
where
\begin{eqnarray} \label{cH}
c_H^\mathrm{(in)} &=& c_1 \frac{\Gamma(1 + 2 i b)}{\Gamma\left( \frac12 - i a + i b \right)} + c_2 \frac{\Gamma(1 - 2 i b)}{\Gamma\left( \frac12 - i a - i b \right)},
\nonumber\\
c_H^\mathrm{(out)} &=& i c_1 \frac{\Gamma(1 + 2 i b)}{\Gamma\left( \frac12 + i a + i b \right)} e^{\pi a - \pi b} + i c_2 \frac{\Gamma(1 - 2 i b)}{\Gamma\left( \frac12 + i a - i b \right)} e^{\pi a + \pi b}.
\end{eqnarray}
Near the horizon, the dominant factor is $e^{\pm z/2}$, and consequently the factor $e^{z/2} \sim e^{i/\rho}$ indicates the ingoing mode while $e^{-z/2} \sim e^{- i/\rho}$ corresponds to the outgoing mode.

The corresponding fluxes of each mode can be computed directly via the formula~(\ref{flux}) as
\begin{eqnarray}
D_B^\mathrm{(in)} = - 2 \omega Q^2 \left( 2 b \, e^{- \pi b} \right) \left| c_B^\mathrm{(in)} \right|^2, &\qquad&  D_H^\mathrm{(in)} = - 2 \omega Q^2 \, e^{\pi a} \left| c_H^\mathrm{(in)} \right|^2,
\nonumber\\
D_B^\mathrm{(out)} = 2 \omega Q^2 \left( 2 b \, e^{\pi b} \right) \left| c_B^\mathrm{(out)} \right|^2, &\qquad&  D_H^\mathrm{(out)} = 2 \omega Q^2 \, e^{- \pi a} \left| c_H^\mathrm{(out)} \right|^2.
\end{eqnarray}

\subsection{Outer Boundary Condition}
The boundary condition for the emission of charged particles is that charged pairs are produced by a strong field region near the horizon and the charges with the same sign as the black hole are emitted to the spatial infinity through electric repulsion. The outer boundary condition imposes $D_B^\mathrm{(in)} = 0$ (i.e. $c_1 = 0$).

The transmitted (outgoing) flux at the boundary is
\begin{eqnarray}
D_B^\mathrm{(out)} = 2 \omega Q^2 \, (2b e^{\pi b}) |c_2|^2,
\end{eqnarray}
where the first factor comes from $z^{1/2-ib}$, and the reflected (ingoing) and the incident (outgoing) fluxes near the horizon are, respectively,
\begin{eqnarray}
D_H^\mathrm{(in)} &=& - 2 \omega Q^2 \, e^{\pi a} \left| \frac{\Gamma(1 - 2 i b)}{\Gamma\left( \frac12 - i a - i b \right)}  \right|^2 |c_2|^2,
\\
D_H^\mathrm{(out)} &=& 2 \omega Q^2 \, e^{\pi a + 2 \pi b} \left| \frac{\Gamma(1 - 2 i b)}{\Gamma\left( \frac12 + i a - i b \right)} \right|^2 |c_2|^2.
\end{eqnarray}
Thus the magnitude squares of the Bogoliubov coefficients are given by
\begin{eqnarray}
|\alpha|^2 &=& \frac{D_\mathrm{incident}}{D_\mathrm{reflected}} = \frac{\left| D_H^\mathrm{(out)} \right|}{\left| D_H^\mathrm{(in)} \right|} = \frac{\cosh (\pi a - \pi b)}{\cosh (\pi a + \pi b)} e^{2 \pi b}, \label{Ealpha}
\\
|\beta|^2 &=& \frac{D_\mathrm{transmitted}}{D_\mathrm{reflected}} = \frac{\left| D_B^\mathrm{(out)} \right|}{\left| D_H^\mathrm{(in)} \right|} = \frac{\sinh (2 \pi b)}{\cosh (\pi a + \pi b)} e^{ \pi b - \pi a}. \label{Ebeta}
\end{eqnarray}
These coefficients satisfy the relation $|\alpha|^2 - |\beta|^2 = 1$ from the quantization of the field. The mean number of charged pairs produced via the Schwinger mechanism from the black hole is $|\beta|^2$. In the limit $q \gg m$, the emission rate is approximately given by the Schwinger formula
\begin{eqnarray} \label{Schw form}
|\beta|^2 \approx e^{- \frac{\pi m^2 Q}{q}} \approx e^{- \frac{\pi m^2 r_H^2}{q Q}}.
\end{eqnarray}
This result verifies the speculation that the spontaneous pair production actually occurs near the horizon of RN black holes. The absorption cross section can also be obtained straightforwardly
\begin{equation}
\sigma_\mathrm{abs} = \frac{D_\mathrm{transmitted}}{D_\mathrm{incident}} = \frac{|\beta|^2}{|\alpha|^2} = e^{- \pi a - \pi b} \frac{\sinh(2 \pi b)}{\cosh(\pi a - \pi b)}.
\end{equation}

\subsection{Inner Boundary Condition}
The inner boundary condition assumes $D_H^\mathrm{(out)} = 0$ (i.e. $c_H^\mathrm{(out)} = 0$), which gives the following relation for two undetermined parameters, $c_1$ and $c_2$, in the general solution
\begin{equation}
c_2 = - c_1 \frac{\Gamma\left( 1 + 2ib \right) \Gamma\left( \frac12 + ia - ib \right)}{\Gamma\left( 1 - 2ib \right) \Gamma\left( \frac12 + ia + ib \right)} e^{-2\pi b},
\end{equation}
or, equivalently leads to
\begin{equation}
c_H^\mathrm{(in)} = c_1 e^{-\pi(a + b)} \frac{\sinh(2 \pi b)}{\cosh(\pi a - \pi b)} \frac{\Gamma\left( 1 + 2ib \right)}{\Gamma\left( \frac12 - ia + ib \right)}.
\end{equation}

The magnitude squares of the Bogoliubov coefficients, after a routine calculation, are given by
\begin{eqnarray}
|\alpha|^2 &=& \frac{D_\mathrm{incident}}{D_\mathrm{reflected}} = \frac{\left| D_B^\mathrm{(in)} \right|}{\left| D_B^\mathrm{(out)} \right|} = \frac{\cosh (\pi a - \pi b)}{\cosh (\pi a + \pi b)} e^{2 \pi b},
\\
|\beta|^2 &=& \frac{D_\mathrm{transmitted}}{D_\mathrm{reflected}} = \frac{\left| D_H^\mathrm{(in)} \right|}{\left| D_B^\mathrm{(out)} \right|} = \frac{\sinh (2 \pi b)}{\cosh (\pi a + \pi b)} e^{ \pi b- \pi a},
\end{eqnarray}
and the absorption cross section is given by
\begin{eqnarray} \label{Esigma}
\sigma_\mathrm{abs} &=& \frac{D_\mathrm{transmitted}}{D_\mathrm{incident}} = \frac{|\beta|^2}{|\alpha|^2} = \frac{\left| D_H^\mathrm{(in)} \right|}{\left| D_B^\mathrm{(in)} \right|} = e^{- \pi a - \pi b} \frac{\sinh(2 \pi b)}{\cosh(\pi a - \pi b)}
\nonumber\\
&=& \frac1{\pi} e^{- \pi a - \pi b} \sinh(2 \pi b) \left| \Gamma\left( \frac12 + ia - ib \right) \right|^2.
\end{eqnarray}

As shown in Sec.~\ref{equivalence}, $|\alpha|^2, |\beta|^2$ and $\sigma_\mathrm{abs}$ are actually identical with those obtained by imposing the outer boundary condition.

\subsection{Quasi-normal Modes}
For the quasi-normal mode analysis, we should impose both the inner and outer boundary conditions at the same time: $D_B^\mathrm{(in)} = 0$ and $D_H^\mathrm{(out)} = 0$. The only possible solution for the first condition comes from $c_1 = 0$. For the second condition, beside the trivial choice $c_2 = 0$, there is another non-trivial condition
\begin{equation}
\frac1{\Gamma\left( \frac12 + ia - ib \right)} = 0,
\end{equation}
which is fulfilled only when
\begin{equation} \label{Eqn}
\frac12 + ia - ib = - N, \quad N = 0, 1, \cdots.
\end{equation}
Since parameters $a$ and $b$ are real and $b < a$, the condition~(\ref{Eqn}) cannot be satisfied, which means that there is always outgoing modes at the black hole horizon in the extremal limit. That is, there are no quasi-normal modes in this limit.

\section{Near-Extremal Black Holes}

\subsection{Solution of Scalar Field}
For the near-extremal case, i.e. $B \ne 0$, the radial equation~(\ref{eom}) has the following solution in terms of the hypergeometric function
\begin{eqnarray}
R(\rho) &=& c_1 (\rho - B)^{- \frac{i}2 (a - \tilde a)} (\rho + B)^{\frac{i}2 (a + \tilde a)} \, F\left( \frac12 + i \tilde a + i b, \frac12 + i \tilde a - i b; 1 - i a + i \tilde a; \frac12 - \frac{\rho}{2B} \right)
\nonumber\\
&+& c_2 (\rho - B)^{\frac{i}2 (a - \tilde a)} (\rho + B)^{\frac{i}2 (a + \tilde a)} \, F\left( \frac12 + i a + i b, \frac12 + i a - i b; 1 + i a - i \tilde a; \frac12 - \frac{\rho}{2B} \right),
\end{eqnarray}
where parameters $a, b$ are in Eq.~(\ref{ab}) and the new parameter $\tilde a$ is defined as\footnote{In terms of the variables of the original RN black holes according to the rescaling~(\ref{rescal})
$$ B = \frac{\sqrt{2Q (M - Q)}}{\epsilon}, \qquad \omega = \frac{w}{\epsilon}, $$
the parameter $\tilde a$ can be expressed as
$$ \tilde a = \frac{w Q^2}{\sqrt{2Q (M - Q)}}, $$
where $w$ denotes the frequency with respect to the RN time coordinate $t$ as $e^{-i w t} = e^{-i \omega \tau}$.
}
\begin{equation}
\tilde a \equiv \frac{\omega Q^2}{B}.
\end{equation}
The frequency (energy) dependence is a significant property of the parameter $\tilde a$ which is essentially related to the Hawking radiation as  a tunneling effect on the black hole horizon.

As the horizon is located at $\rho = B$, by the property~(\ref{limF0}), the solution at the near horizon region reduces to
\begin{eqnarray}
R_H(\rho) &=& c_H^\mathrm{(in)} (\rho - B)^{- \frac{i}2 (\tilde a - a)} (\rho + B)^{\frac{i}2 (\tilde a + a)} + c_H^\mathrm{(out)} (\rho - B)^{\frac{i}2 (\tilde a - a)} (\rho + B)^{\frac{i}2 (\tilde a + a)}
\nonumber\\
&\approx& c_H^\mathrm{(in)} (2B)^{\frac{i}2 (\tilde a + a)} (\rho - B)^{- \frac{i}2 (\tilde a - a)} + c_H^\mathrm{(out)} (2B)^{\frac{i}2 (\tilde a + a)} (\rho - B)^{\frac{i}2 (\tilde a - a)},
\end{eqnarray}
where
\begin{equation}
c_H^\mathrm{(in)} = c_2, \qquad c_H^\mathrm{(out)} = c_1.
\end{equation}
On the other hand, at the asymptotic boundary, by the properties~(\ref{relF3}, \ref{limF0}), the solution behaves as
\begin{eqnarray} \label{RBasy}
R_B(\rho) &=& c_B^\mathrm{(in)} (\rho - B)^{-\frac12 - \frac{i}2 (\tilde a + a) - i b} (\rho + B)^{\frac{i}2 (\tilde a + a)} + c_B^\mathrm{(out)} (\rho - B)^{-\frac12 - \frac{i}2 (\tilde a + a) + i b} (\rho + B)^{\frac{i}2 (\tilde a + a)}
\nonumber\\
&\approx& c_B^\mathrm{(in)} \rho^{-\frac12 - i b} + c_B^\mathrm{(out)} \rho^{-\frac12 + i b},
\end{eqnarray}
where
\begin{eqnarray}
c_B^\mathrm{(in)} &=& c_1 (2 B)^{\frac12 + i \tilde a + i b} \frac{\Gamma(1 - i a + i \tilde a) \Gamma(- 2 i b)}{\Gamma\left( \frac12 - i a - i b \right) \Gamma\left( \frac12 + i \tilde a - i b \right)} + c_2 (2 B)^{\frac12 + i a + i b} \frac{\Gamma(1 + i a - i \tilde a) \Gamma(- 2 i b)}{\Gamma\left( \frac12 + i a - i b \right) \Gamma\left( \frac12 - i \tilde a - i b \right)},
\\
c_B^\mathrm{(out)} &=& c_1 (2 B)^{\frac12 + i \tilde a - i b} \frac{\Gamma(1 - i a + i \tilde a) \Gamma( 2 i b)}{\Gamma\left( \frac12 - i a + i b \right) \Gamma\left( \frac12 + i \tilde a + i b \right)} + c_2 (2 B)^{\frac12 + i a - i b} \frac{\Gamma(1 + i a - i \tilde a) \Gamma( 2 i b)}{\Gamma\left( \frac12 + i a + i b \right) \Gamma\left( \frac12 - i \tilde a + i b \right)}.
\end{eqnarray}
Using the approximate solutions near the both inner and outer boundaries, the corresponding fluxes of each mode can be obtained
\begin{eqnarray}
D_B^\mathrm{(in)} = - 2 b \left| c_B^\mathrm{(in)} \right|^2, &\qquad&  D_H^\mathrm{(in)} = - 2 B (\tilde a - a) \left| c_H^\mathrm{(in)} \right|^2,
\nonumber\\
D_B^\mathrm{(out)} = 2 b \left| c_B^\mathrm{(out)} \right|^2, &\qquad&  D_H^\mathrm{(out)} = 2 B (\tilde a - a) \left| c_H^\mathrm{(out)} \right|^2.
\end{eqnarray}

\subsection{Outer Boundary Condition}
In this subsection, we compute the magnitude squares of the Bogoliubov coefficients and the absorption cross section by imposing the outer boundary condition, i.e. $c_B^\mathrm{(in)} = 0$. This condition relates the parameters $c_1$ and $c_2$
\begin{equation}
c_1 = - c_2 (2B)^{i(a - \tilde a)} \frac{\Gamma(1 + i a - i \tilde a) \Gamma\left( \frac12 - i a - i b \right) \Gamma\left( \frac12 + i \tilde a - i b \right)}{\Gamma(1 - i a + i \tilde a) \Gamma\left( \frac12 + i a - i b \right) \Gamma\left( \frac12 - i \tilde a - i b \right)},
\end{equation}
which leads to
\begin{equation}
c_B^\mathrm{(out)} = - c_2 (2 B)^{\frac12 + i a - i b} \frac{\sinh(2\pi b) \sinh(\pi \tilde a - \pi a)}{\cosh(\pi a + \pi b) \cosh(\pi \tilde a - \pi b)} \frac{\Gamma(1 + i a - i \tilde a) \Gamma( 2 i b)}{\Gamma\left( \frac12 + i a + i b \right) \Gamma\left( \frac12 - i \tilde a + i b \right)}.
\end{equation}

The magnitude squares of the Bogoliubov coefficients are
\begin{eqnarray}
|\alpha|^2 &=& \frac{D_\mathrm{incident}}{D_\mathrm{reflected}} = \frac{\left| D_H^\mathrm{(out)} \right|}{\left| D_H^\mathrm{(in)} \right|} = \frac{\cosh (\pi a - \pi b) \cosh (\pi \tilde a + \pi b)}{\cosh (\pi a + \pi b) \cosh (\pi \tilde a - \pi b)}, \label{Nalpha}
\\
|\beta|^2 &=& \frac{D_\mathrm{transmitted}}{D_\mathrm{reflected}} = \frac{\left| D_B^\mathrm{(out)} \right|}{\left| D_H^\mathrm{(in)} \right|} = \frac{\sinh (2 \pi b) \sinh (\pi \tilde a - \pi a )}{\cosh (\pi a + \pi b) \cosh (\pi \tilde a - \pi b)}. \label{Nbeta}
\end{eqnarray}
As a consequence of the flux conservation, these coefficients satisfy the Bogoliubov relation $|\alpha|^2 - |\beta|^2 = 1$. In the limit $B \to 0$ ($\tilde a \to \infty$), the results reduce to the extremal case~(\ref{Ealpha}, \ref{Ebeta}). Moreover, the leading term of $|\beta|^2$ leads to the Schwinger formula~(\ref{Schw form}). The absorption cross section is
\begin{equation}
\sigma_\mathrm{abs} = \frac{D_\mathrm{transmitted}}{D_\mathrm{incident}} = \frac{|\beta|^2}{|\alpha|^2} = \frac{\sinh (2 \pi b) \sinh (\pi \tilde a - \pi a )}{\cosh (\pi a - \pi b) \cosh (\pi \tilde a + \pi b)}.
\end{equation}
Again, it reduces to~(\ref{Esigma}) in the limit $\tilde a \to \infty$.

It has been pointed out that the spontaneous pair production in the extremal black hole is generated completely by the Schwinger mechanism in which the charged particles with the same charge as that of the black hole are repulsed to infinity by the constant electric field. One may naively expect to distinguish the Hawking radiation from the Schwinger mechanism by comparing the production rate in the near extremal case (Schwinger + Hawking) with the production rate in the extremal limit (pure Schwinger). A remarkable fact is observed from this comparison: the mean numbers $|\beta|^2$ in the extremal~(\ref{Ebeta}) and in the near extremal~(\ref{Nbeta}) cases have the ratio
\begin{equation}
\frac{|\beta(B = 0)|^2}{|\beta(B \ne 0)|^2} = \frac{\cosh(\pi \tilde a - \pi b)}{\sinh(\pi \tilde a - \pi a)} e^{\pi b - \pi a} = \frac{1 + e^{2 \pi (b - \tilde a)}}{1 - e^{2 \pi (a - \tilde a)}} \ge 1.
\end{equation}
This ratio indicates that the production rate in the extremal limit (Schwinger) is greater than the production rate in the near extremal case (Schwinger + Hawking). A major physical interaction dominates here: as the geometry changes from the extremal to near extremal black holes, the increasing attractive gravitational force will reduce the electromagnetic repulsive force for the Schwinger mechanism. Therefore, the production rate of the Schwinger mechanism is suppressed faster than the increasing part from the Hawking thermal radiation. Moreover, such kind of interaction generically prohibits one distinguishing the Schwinger mechanism from the Hawking radiation.

\subsection{Inner Boundary Condition}
For the comparison, we repeat the computation by imposing the inner boundary condition, namely $c_H^\mathrm{(out)} = c_1 = 0$. The magnitude squares of the Bogoliubov coefficients are
\begin{eqnarray}
|\alpha|^2 &=& \frac{D_\mathrm{incident}}{D_\mathrm{reflected}} = \frac{\left| D_B^\mathrm{(in)} \right|}{\left| D_B^\mathrm{(out)} \right|} = \frac{\cosh (\pi a - \pi b) \cosh (\pi \tilde a + \pi b)}{\cosh (\pi a + \pi b) \cosh (\pi \tilde a - \pi b)},
\\
|\beta|^2 &=& \frac{D_\mathrm{transmitted}}{D_\mathrm{reflected}} = \frac{\left| D_H^\mathrm{(in)} \right|}{\left| D_B^\mathrm{(out)} \right|} = \frac{\sinh (2 \pi b) \sinh (\pi \tilde a - \pi a )}{\cosh (\pi a + \pi b) \cosh (\pi \tilde a - \pi b)},
\end{eqnarray}
and the absorption cross section is
\begin{eqnarray} \label{Nsigma}
\sigma_\mathrm{abs} &=& \frac{D_\mathrm{transmitted}}{D_\mathrm{incident}} = \frac{\left| D_H^\mathrm{(in)} \right|}{\left| D_B^\mathrm{(in)} \right|} = \frac{\sinh (2 \pi b) \sinh (\pi \tilde a - \pi a )}{\cosh (\pi a - \pi b) \cosh (\pi \tilde a + \pi b)}
\nonumber\\
&=& \frac1{\pi^2} \sinh(2 \pi b) \sinh (\pi \tilde a - \pi a ) \left| \Gamma\left( \frac12 + ia - ib \right) \right|^2 \left| \Gamma\left( \frac12 + i \tilde a + ib \right) \right|^2.
\end{eqnarray}
These quantities are the same as those from the outer boundary condition, as expected from Sec.~\ref{equivalence}.

\subsection{Quasi-normal Modes}
The quasi-normal mode boundary condition, $D_B^\mathrm{(in)} = 0$ and $D_H^\mathrm{(out)} = 0$, requires trivial solution $c_1 = c_2 = 0$ except some special discrete values of parameters, i.e. quasi-normal modes, determined by the condition
\begin{equation}
\frac1{\Gamma\left( \frac12 + ia - ib \right) \Gamma\left( \frac12
-i \tilde a - ib \right)} = 0.
\end{equation}
Therefore, beside the condition given in Eq.~(\ref{Eqn}) for the extremal limit, there is an additional possibility for the near extremal black holes
\begin{equation} \label{Nqn}
\frac12 - i\tilde a - ib = - N, \quad N = 0, 1, \cdots.
\end{equation}
This condition gives the quasi-normal mode frequencies
\begin{equation}
\omega = -\frac{b B}{Q^2} - i \left( \frac12 + N \right) \frac{B}{Q^2}.
\end{equation}
Since the parameter $B > 0$, the quasi-normal mode boundary condition thus requires the appearance of the negative energy states.

\section{Holographic Dual Description}
The spontaneous production in Secs. IV and V corresponds to the instability of the probe charged scalar field in the AdS space. One expects that the charged scalar field should be dual to an ``unstable'' operator in the boundary conformal field theory (CFT) according to the holographic principle. Previous studies on the RN/CFT correspondence~\cite{Chen:2010bsa, Chen:2010as, Chen:2011gz} shows that the 4D (near) extremal RN black hole consists of an AdS$_3 = $ AdS$_2 \times S^1$ structure, where the AdS$_2$ comes from the near horizon geometry and the U(1) gauge field plays the role of the $S^1$ bundle. The central charges and temperatures of the dual two-dimensional CFT have been determined as
\begin{equation}
c_L = c_R = \frac{6 Q^3}{\ell}, \qquad T_L = \frac{\ell}{2\pi Q}, \qquad T_R = \frac{\ell B}{\pi Q^2},
\end{equation}
where $\ell$ is a free parameter which can be interpreted as a measure of the U(1) bundle. In addition, from the asymptotic form of the solution, either~(\ref{Rasy}) or~(\ref{RBasy}), we can see that unlike the conventional situation of massive probe fields in the AdS/CFT correspondence~\cite{Maldacena:1997re, Gubser:1998bc, Witten:1998qj}, the operator and the source on the boundary are of the same magnitude, namely, the conformal weights of the dual operator are
\begin{equation}
h_L = h_R = \frac12 \pm ib,
\end{equation}
without loss of generality, we choose $h_L = h_R = 1/2 + ib$ below. The complex conformal weight means that the dual operator is unstable.

To compare the results calculated from the gravity side with those from the CFT's, recall that the general expression for a two-point function $G(\sigma_+, \sigma_-)$ in 2D CFT is
\begin{equation}
G(\sigma_+, \sigma_-) = \langle \phi(\sigma_+) \, \phi(\sigma_-) \rangle = (-)^{h_L + h_R} \left( \frac{\pi T_L}{\sinh(\pi T_L \sigma_+)} \right)^{2h_L} \left( \frac{\pi T_R}{\sinh(\pi T_R \sigma_-)} \right)^{2h_R} e^{i q_L \Omega_L \sigma_+ + i q_R \Omega_R \sigma_-},
\end{equation}
where ($q_L, q_R$) and ($\Omega_L, \Omega_R$) are the charges and chemical potentials of the left- and right- hands operators, respectively. Consequently, the absorption cross section of the operator $\phi$ is
\begin{eqnarray} \label{abcft}
\sigma_{abs} &\sim & \frac{(2 \pi T_L)^{2h_L-1}}{\Gamma(2 h_L)} \frac{(2 \pi T_R)^{2 h_R-1}}{\Gamma(2 h_R)} \sinh\left( \frac{\omega_L - q_L \Omega_L}{2 T_L} + \frac{\omega_R - q_R \Omega_R}{2 T_R} \right)
\nonumber\\
&& \times \left| \Gamma\left( h_L + i \frac{\omega_L - q_L \Omega_L}{2 \pi T_L} \right) \right|^2 \left| \Gamma\left( h_R + i \frac{\omega_R - q_R \Omega_R}{2 \pi T_R} \right) \right|^2.
\end{eqnarray}
Further identifying the first law of thermodynamics of the black hole with that of the dual 2D CFT, i.e. $\delta S_{\rm BH} = \delta S_{\rm CFT}$, we have
\begin{equation}
\frac{\delta M}{T_H} - \frac{\Omega_H \delta Q}{T_H} = \frac{\tilde{\omega}_L}{T_L} + \frac{\tilde{\omega}_R}{T_R},
\end{equation}
where the black hole Hawking temperature and chemical potential are $T_H = \frac{B}{2\pi Q^2}, \, \Omega_H = A_\tau(B) = - B/Q$ and $\tilde{\omega}_L = \omega_L - q_L \Omega_L, \, \tilde{\omega}_R = \omega_R - q_R \Omega_R$. Together with the identifications $\delta M = \omega$ and $\delta Q = -q$ (the minus sign corresponds to the convention ``$-q$'' in the operator $D_\alpha$ for the dynamics of the charged scalar field), we can then determine that
\begin{equation}
\tilde{\omega}_L = - q \ell \quad {\rm and} \quad \tilde{\omega}_R = 2 \omega \ell.
\end{equation}
One can see that the absorption cross section~(\ref{Nsigma}) agrees with the CFT's result~(\ref{abcft}) only up to some numerical factors. The dual description for the production rate $|\beta|^2$ can be understood through the relation~(\ref{betasigma}).

\section{Conclusion}
We have studied the spontaneous pair production of charge scalar field for the RN black holes in the extremal and near extremal limits. The charged particle pairs are produced near the horizon region of the black holes that has a specific spacetime structure AdS$_2 \times S^2$ with a constant electric field. The Bogoliubov coefficients and the absorption cross section have been computed from the ratios of fluxes through the outer and inner boundaries by imposing two suitable boundary conditions. Indeed these two boundary conditions are equivalent and admit to study the spontaneous production as discussed in Section~\ref{SBC}. The Bogoliubov coefficients provide the information of the vacuum polarization and the vacuum persistence on the horizon. The model in this paper is one of the few examples that we exactly know the Bogoliubov coefficients. In particular, the explicit expressions of the vacuum persistence amplitude $|\alpha|^2$, the mean number of pairs $|\beta|^2$, and the absorption cross section $\sigma_\mathrm{abs}$ have been obtained. Moreover, the holographic description of the pair production in which the emitted particles are unstable
tachyon modes has also been discussed.

By comparing the results in the extremal and near extremal cases, it is shown that the production rate is suppressed when the black hole temperature is turned on. This is a consequence of the attractive gravitational force that reduces the repulsive electromagnetic force for the Schwinger mechanism and thereby suppresses the corresponding production rate. On the contrary, a repulsive gravitational force, for example, in a de Sitter space with positive cosmological constant, enhances the Schwinger pair production. In virtue of such kind of interactions, the effects of the Schwinger mechanism and the Hawking radiation generically cannot be distinguished by imposing different boundary conditions.

\section*{Acknowledgement}
The authors thank Akihiro Ishibashi, Nobuyoshi Ohta and Don N. Page for useful discussions. CMC was supported by the National Science Council of the R.O.C. under the grant NSC 99-2112-M-008-005-MY3 and in part by the National Center of Theoretical Sciences (NCTS). The work of SPK was supported by the National Research Foundation (NRF) Grant funded by the Korean Government (MEST)(2010-0016-422). JRS was supported by the NSFC Grant No. 11147190.

\begin{appendix}
\section{Useful Properties of Special Functions}
In this Appendix, we list useful properties of some special functions that are used in our computations. The details may be found, for instance, in~\cite{GR94}.

The Wittaker's equation
\begin{equation}
\frac{d^2}{dz^2} w(z) + \left( - \frac14 + \frac{\kappa}{z} + \frac{\frac14 - \mu^2}{z^2} \right) w(z) = 0,
\end{equation}
has the solutions, which are called the Whittaker functions
\begin{eqnarray}
M_{\kappa, \mu}(z) &=& e^{-\frac{z}2} z^{\frac12 + \mu} F\left( \frac12 + \mu - \kappa, 1 + 2 \mu, z \right),
\\
W_{\kappa, \mu}(z) &=& e^{-\frac{z}2} z^{\frac12 + \mu} U\left( \frac12 + \mu - \kappa, 1 + 2 \mu, z \right).
\end{eqnarray}
In the case of non-integer $2 \mu$, the Whittaker functions have following relations
\begin{eqnarray}
W_{\kappa, \mu}(z) &=& \frac{\Gamma(-2 \mu)}{\Gamma\left( \frac12 - \mu - \kappa \right)} M_{\kappa, \mu}(z) + \frac{\Gamma(2 \mu)}{\Gamma\left( \frac12 + \mu - \kappa \right)} M_{\kappa, -\mu}(z), \qquad \arg z < \frac32 \pi, \label{relW}
\\
M_{\kappa, \mu}(z) &=& \frac{\Gamma (1 + 2\mu)}{\Gamma\left( \frac12 + \mu - \kappa \right)} e^{- i \pi \kappa} W_{-\kappa, \mu} (e^{-i \pi} z) + \frac{\Gamma(1 + 2\mu)}{\Gamma\left( \frac12 + \mu + \kappa \right)} e^{i \pi \left( \frac12 + \mu - \kappa \right)} W_{\kappa, \mu}(z), \quad -\frac12 \pi < \arg z < \frac32 \pi. \label{relM}
\end{eqnarray}
Moreover, these two special functions have the following asymptotic forms
\begin{equation} \label{limMW}
\lim_{|z| \to 0} M_{\kappa, \mu}(z) \to e^{-\frac{z}2} z^{\frac12 + \mu}, \qquad \lim_{|z| \to \infty} W_{\kappa, \mu}(z) \to e^{-\frac{z}2} z^\kappa.
\end{equation}

We also used the transformation formula of the hypergeometric function,
\begin{eqnarray}
F(a, b; c; z)
&=& \frac{\Gamma(c) \Gamma(b - a)}{\Gamma(b) \Gamma(c - a)} (-z)^{-a} F\left( a, 1 - c + a; 1 - b + a; \frac1{z} \right)
\nonumber\\
&& + \frac{\Gamma(c) \Gamma(a - b)}{\Gamma(a) \Gamma(c - b)} (-z)^{-b} F\left( b, 1 - c + b; 1 - a + b; \frac1{z} \right) \qquad \left( |\arg (-z)| < \pi \right). \label{relF3}
\end{eqnarray}
and a special value
\begin{equation} \label{limF0}
F(a, b; c; 0) = 1.
\end{equation}


Finally, we have used the particular relations for the Gamma function
\begin{equation}
\left| \Gamma\left( \frac12 + iy \right) \right|^2 = \frac{\pi}{\cosh \pi y}, \qquad \left| \Gamma\left(1 + iy \right) \right|^2 = \frac{\pi y}{\sinh \pi y}, \qquad \left| \Gamma\left( iy \right) \right|^2 = \frac{\pi}{y \sinh\pi y},
\end{equation}
and worked in the Riemann sheet $- 3\pi/2 < \arg z < \pi/2$ for analytical continuations
\begin{equation}
i = e^{i \pi/2}, \qquad -1 = e^{- i \pi}, \qquad -i = e^{- i \pi/2}.
\end{equation}

\section{Hamilton-Jacobi Approach to Pair Production}
The $|\beta|^2$ in~(\ref{Nbeta}) can be expressed as
\begin{equation}
|\beta|^2 = \frac{\sinh (2 \pi b)}{\cosh (\pi a + \pi b)} \frac{\sinh (\pi \tilde a - \pi a )}{\cosh (\pi \tilde a - \pi b)} = e^{2\pi b - 2\pi a} \Bigl( \frac{1 - e^{-4\pi b}}{1 + e^{-2\pi a - 2\pi b}} \frac{1 - e^{-2\pi \tilde a + 2\pi a}}{1 + e^{-2\pi \tilde a + 2\pi b}}
\Bigr). \label{app-B-1}
\end{equation}
In the limit $a \sim b \gg 1$ $(q \gg m)$ considered in literature, the leading term
\begin{equation}
e^{2\pi b - 2\pi a} \approx e^{-\pi \left( \frac{m^2 Q}{q} + \frac{(l+1/2)^2}{q Q^2}\right)},
\end{equation}
and other terms in parenthesis~(\ref{app-B-1}) include corrections to the Schwinger formula.

\begin{figure}[ht]
\includegraphics[width=7cm]{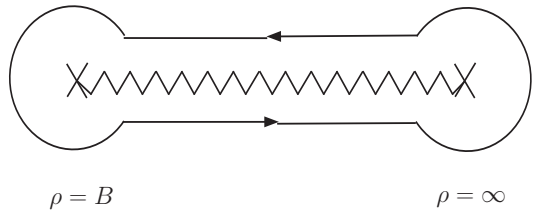}
\caption{The contour integral for pair production in phase-integral method.} \label{F3}
\end{figure}

In order to understand the leading term in~(\ref{app-B-1}), we use the Hamilton-Jacobi approach to pair production
via quantum tunneling. In the phase-integral method the Hamilton-Jacobi approximation $R(\rho) = e^{i S(\rho)}$ to~(\ref{EqR}) has the imaginary part~\cite{Kim:2007ab}
\begin{eqnarray}
2 \, {\rm Im} S &=& -i \oint d \rho \sqrt{\frac{(q \rho - \omega Q)^2 Q^2}{(\rho^2 - B^2)^2} - \frac{m^2 Q^2 + (l + 1/2)^2}{\rho^2 - B^2} }
\nonumber\\
&=& -i b \oint d \rho \frac{\sqrt{(\rho - \rho_+) (\rho - \rho_-)}}{(\rho - B) (\rho + B)},
\end{eqnarray}
where
\begin{eqnarray}
\rho_{\pm} = \frac{q \omega Q^3}{b^2} \pm \frac{q \omega Q^3}{b^2} \sqrt{ 1- \frac{b^2}{q^2 Q^2} - \frac{m^2 Q^2 + (l + 1/2)^2}{q^2 \omega^2 Q^6} b^2 B^2}.
\end{eqnarray}
The contour integral excludes the branch cut in Fig.~\ref{F3} and poles are located at $\rho = B$ and $\rho = \infty$.

Firstly, in the case of extremal black hole $(B = 0)$ the contour integral exterior to the branch cut yields
\begin{eqnarray}
2 {\rm Im} S = 2 \pi (a - b), \label{app-B-im S}
\end{eqnarray}
where $2 \pi a$ comes from the small-$\rho$ expansion and $2 \pi b$ from the large-$\rho$ expansion. Thus the pair production is approximately given by $|\beta|^2 \approx e^{- 2 {\rm Im} S}$, which accounts for the leading term in~(\ref{app-B-1}). We compare the instanton action~(\ref{app-B-im S}) with Eq.~(39) of Ref.~\cite{Khriplovich:1999gm}, which mostly contributes to~(\ref{app-B-1}) near the peak of the barrier for $\omega = q$. In Ref.~\cite{Gabriel:2000mg} the radial part~(\ref{EqR}) of the KG equation is approximated by an inverted parabola, which is equivalent to the motion in a constant electric field with $E = \omega^2/(q^2 Q)$ and leads to pair production $e^{- \pi m^2 qQ/\omega^2} = e^{- \pi m^2 Q/q}$ for $\omega = q$. Secondly, in the case of near extremal black holes, the large-$\rho$ and small-$\rho$ expansions contribute
\begin{eqnarray}
2 \, {\rm Im} S = \pi (\tilde{a} - a) - 2 \pi b.
\end{eqnarray}

\end{appendix}


\end{document}